\documentclass[	prl,eclepsf,showpacs,amsmath,amssymb,twocolumn]{revtex4}

\usepackage{graphicx}
\usepackage{amsmath}
\usepackage{bm}
\usepackage{multirow}
\usepackage{comment}
\graphicspath{{./Figs/}}
\usepackage{color}
\input{colordvi.tex}
\newcommand{\simgt}{\lower.5ex\hbox{$\; \buildrel > \over \sim \;$}}
\newcommand{\simlt}{\lower.5ex\hbox{$\; \buildrel < \over \sim \;$}}

\def\mpl{{M_{\rm Pl}}}
\def\rvir{{r_{\rm vir}}}
\def\mvir{{M_{\rm vir}}}

\begin{document}

\title{Gas density profile in dark matter halo in chameleon cosmology}

\author{Ayumu Terukina}
\email[Email:]{telkina"at"theo.phys.sci.hiroshima-u.ac.jp}

\author{Kazuhiro Yamamoto}
\email[Email:]{kazuhiro"at"hiroshima-u.ac.jp}

\address{
Department of Physical Science, Hiroshima University,
Higashi-Hiroshima 739-8526,~Japan}
\date{\today}

\begin{abstract}
We investigate the gas density, temperature, and pressure profiles in 
a dark matter halo under the influence of the chameleon force. 
We solve the hydrostatic equilibrium equation for the gas coupled with the 
chameleon field in an analytic manner, using an approximate solution for the 
chameleon field equation with the source term, with a generalized 
Navarro-Frenk-White universal density profile. 
We find that the gas distribution becomes compact because a larger pressure 
gradient is necessary due to the additional chameleon force.
By confronting the theoretical prediction with the data of the temperature 
profile of the Hydra A cluster according to Suzaku x-ray  observations out to the virial 
radius, we demonstrate that a useful constraint on a model parameter can be 
obtained depending on the value of the coupling constant. 
For example, the upper bound of the background value of chameleon field,
$\phi_\infty<10^{-4}\mpl$, is obtained in the case $\beta=1$, where $\beta$
is the coupling constant between the chameleon field and the matter, and $\mpl$ is the Planck mass.
However, the error of the present data is not so small that we obtain a useful constraint
in the case $\beta=1/\sqrt{6}$, which corresponds to an $f(R)$ model.

%Hewever, no useful constarint was obtained for an $f(R)$ model
%from the present X-ray data of the Hydra A cluster. 

%To constrain an $f(R)$ model, we need observe smaller mass cluster???

%For a smaller mass halo, the gas distribution could be much 
%smaller than the virial radius, which put a constraint on the chameleon field configuration. 
%We also discuss about an application of our results to an $f(R)$ 
%modified gravity model, which may put a tight constraint on the model 
%parameter of $|f_{R0}|\simlt 10^{-5}$. 
\end{abstract}

\maketitle
%%%%%%%%%%%%%%%%%%%%%%%%%%%%%%%%%%%%%%%%%%%%%%%%%%%%%%%%%%%%%%%%%%%%%%%%%%%%%%%%%

%\newpage
%---------------------------------------------------------------
%\section
%{\it Introduction}
%--------------------------------------------------------------\-
The accelerated expansion of the Universe is one of the most 
fundamental mysteries in basic science.
Long-distance modification of the gravity theory is a challenging 
approach to this problem. 
However, any gravity theory must pass the stringent constraints 
from the Solar System.
The chameleon mechanism is a noble mechanism for screening a scalar degree of 
freedom which appears in a class of modified gravity models, depending 
on the density of matter in the local environment \cite{Chameleonmechanism,Chameleonmechanism2}. 
Newtonian gravity is recovered in a high-density region, thereby evading the Solar 
System constraints.
Recently, it has been pointed out that the modification of gravity might 
be detected using halos of galaxies and galaxy clusters,  because 
the screening mechanism could not be complete in their outer regions  
~\cite{dynamicalmass,Lam,Ki,clusterdensityprofile,Yano,NarikawaYamamoto}.

In the present paper, we focus on the gas distribution in a dark matter halo 
under the influence of the chameleon force. 
In Ref.~\cite{Chameleongravity}, the authors found an analytic
solution of the chameleon field, assuming the matter distribution of the 
Navarro-Frenk-White 
(NFW) universal density profile~\cite{NFWProfile} (c.f. Ref.~\cite{Panpanich}). 
Utilizing their analytic method, we investigate the gas density, 
temperature, and pressure profiles under the influence of the chameleon 
force. 
We find that the chameleon force significantly influences the gas distribution.
We also demonstrate a useful constraint on the chameleon gravity model from 
confronting the theoretical temperature profile with x-ray observations 
of a cluster of galaxies. 

%We demonstrate how these observations may put constraints on the
%chameleon gravity model as well as the $f(R)$ model.

%---------------------------------------------------------------
%\section
%{\it Gas density profile}~
%---------------------------------------------------------------
The chameleon field equation for a quasistatic system in the Einstein frame is
%%%%
\begin{equation}\label{1.1.2}
\nabla^2\phi=V_{,\phi}+\frac{\beta}{\mpl}\rho e^{\beta\phi/\mpl},
\end{equation}
%%%%
where $V$ is the potential, $\rho$ is the matter density, 
$\beta$ is the coupling constant, and we have defined the 
reduced Planck mass by $\mpl^2=1/(8\pi G)$ with the gravitational 
constant $G$.
Here, we assume $V(\phi)=\Lambda^{4+n}/\phi^n$, where $\Lambda$ is 
the mass dimension parameter and $n$ is the dimensionless parameter. 
We also assume $\beta\phi/\mpl\ll1$.
The coupling between the scalar field and the matter density
is the key for the chameleon mechanism, as we see below.

We follow the analytic method in Ref.~\cite{Chameleongravity} to find a solution 
for Eq.~(\ref{1.1.2}). In the present paper, we assume the generalized NFW density 
profile
%%%%
$%\begin{equation}\label{2.1.1}
\rho(x)={\rho_s}/{x(1+x)^b}$
%\end{equation}
%%%%
with $x={r}/{r_s}$, where $\rho_s$ and $r_s$ are the characteristic density 
and scale of a halo, respectively, and $b$ is a parameter. 
The NFW density profile is the case $b=2$.
The mass within the radius $x$ of the halo is given by
%%%%
%\begin{eqnarray}\label{2.1.6_1}
$M(x)=4\pi r_s^3 \int_0^x dxx^2\rho(x)$.
%with $m(x)=\ln(1+x)-{x}/(1+x)$.
%\end{eqnarray}
%%%%
Instead of the parameters $r_s$ and $\rho_s$, 
we introduce the virial mass $\mvir$ and the concentration parameter $c$, 
which are defined by $\mvir= {(4\pi/3)\rvir^3\Delta_c\bar\rho_c}$ and $c= \rvir/r_s$,
%%%%
%\begin{equation}\label{2.1.4}
%$\rvir={\mvir}^{1/3}/[{(4\pi/3)\Delta_c\bar\rho_c}]^{1/3}$,
%\end{equation}
%%%%
where $\Delta_c$ is the ratio of the spherical overdensity 
$\bar\rho(<c)$ within the virial radius $\rvir$ to the critical 
density of the universe $\bar\rho_c$; i.e., 
$\Delta_c=\bar\rho(<c)/\bar\rho_c$, for which we adopt $\Delta_c=100$
in a spherical collapse model.

The analytic solution for Eq.~(\ref{1.1.2}) is obtained by matching 
the interior solution $\phi_{\rm int}$ and the exterior solution $\phi_{\rm out}$, 
where $\phi_{\rm int}$ is given by solving Ep.~(\ref{1.1.2}) 
while neglecting the term on the left-hand-side, while $\phi_{\rm out}$ is given 
by neglecting the first term on the right-hand-side of Eq.~(\ref{1.1.2}). 
Then, we find
\begin{eqnarray}
\phi(x)=\left\{
\begin{array}{l}
\phi_s[x(1+x)^b]^{1/(n+1)}\equiv\phi_{\rm int},~~~~~~{\rm}~x<x_c,\\
-B\dfrac{1-(1+x)^{2-b}}{(b-2)x}
-\dfrac{C}{x}+\phi_\infty\equiv\phi_{\rm out},\\
~~~~~~~~~~~~~~~~~~~~~~~~~~~~~~~~~~~~~~~~~~{\rm}~x>x_c,
\end{array}
\right.
\end{eqnarray}
where we have defined 
$\phi_s=\left({n\Lambda^{n+4}\mpl}/{\beta\rho_s}\right)^{1/(n+1)}$, 
$B={\beta\rho_sr_s^2}/{\mpl}$, and $C$ and $x_c$ are determined 
by solving  the matching conditions at $x=x_c$:
%$C=-B\ln(1+x_c)+\phi_\infty x_c-\phi_s[x_c(1+x_c)^2]^{1/(n+1)}x_c$
%and 
\begin{eqnarray}
&&C=B{(1+x_c)^{2-b}-1\over b-2}+\phi_\infty x_c~~~~~~~~~~~~\nonumber\\
&&~~~~~~~~~~~~~~~~-\phi_s[x_c(1+x_c)^b]^{1/(n+1)}x_c,
\label{defC}
\\
&&\phi_\infty-B(1+x_c)^{1-b}\nonumber\\
&&~~=\phi_s(x_c(1+x_c)^b)^{1/(n+1)}
\left(1+\frac{(1+b)x_c+1}{(n+1)(1+x_c)}\right).
~~~~~~
\label{connection}
\end{eqnarray}
%%%%
%\begin{equation}\label{connection}
%{\phi_s}\left[x_c(1+x_c)^2\right]^{\frac{1}{n+1}}=(n+1)
%\frac{(1+x_c)\phi_\infty-B}{(n+4)x_c+n+2}, 
%\end{equation}
The validity of the analytic solution is demonstrated for the case $b=2$
in Refs.~\cite{Chameleongravity,Ki}.

%%%%
Note that $\phi_s$ is the typical value of the chameleon field
in the interior region, where the chameleon mechanism works. 
This means that $\phi_s\ll \phi_\infty$, because the chameleon 
mechanism screens the chameleon field.
With this fact, Eqs.~(\ref{defC}) and (\ref{connection})  
are approximated as
%$C\simeq-B\ln(1+x_c)+\phi_\infty x_c$.
%{\color{red}
\begin{eqnarray}
&&C\simeq B((1+x_c)^{2-b}-1)/(b-2)+\phi_\infty x_c,\\
&&\phi_\infty-B(1+x_c)^{1-b}\simeq 0,
%{(1+x_c)\phi_\infty-B}\simeq0.
\label{connection2}
\end{eqnarray}
since we are considering the case $x_c={\cal O}(1)$. 
Hence, the scalar field in the exterior region is 
independent of $n$ and $\Lambda$. 
This is important because the constraint we obtain 
becomes independent of $n$ and $\Lambda$. 

In our modeling of the gas distribution in a dark matter halo, 
we make a few assumptions for simplicity. First, we assume that the dark matter 
dominates the dark halo described by the generalized NFW density profile.
The halo density profile could be affected by the modification 
of the gravity theory; however, we assume the same profile, since its 
validity is 
partially supported by $N$-body simulations for the Dvali-Gabadadze-Porrati model and $f(R)$ 
model~\cite{dynamicalmass,Zhao11,Lam,Ki}.
Recently, Ref.~\cite{Ki} confirmed this validity for an $f(R)$ model, 
and argued a qualitative explanation for the validity.
Second, we assume that the baryon density is negligible in the dark matter 
halo, which allows us to neglect its effect on the scalar field equation.
The effect of the baryon component is discussed in 
Ref.~\cite{Universalgasdensity}. The validity of this assumption 
is also supported by the recent measurements of the density
profile of a cluster halo through gravitational lensing, 
which show that the NFW profile fits the data well 
\cite{Umetsu,Oguri}. 
Third, we assume that the scalar field is coupled with the baryon component
as well as the dark matter component. For example, in an $f(R)$ model, 
the chameleon force is coupled with both the dark matter and the baryon 
components. 

Within Newtonian gravity, a useful model of the gas density profile 
is considered in Refs.~\cite{Universalgasdensity,Universalgasdensity2}.
By assuming hydrostatic equilibrium between the gas pressure 
and the gravitational force from the dark matter with the NFW density profile, 
the universal gas density, temperature and pressure profiles are derived. 
We follow the method of Refs.~\cite{Universalgasdensity,Universalgasdensity2}, 
but taking into account the chameleon force as well as the gravitational force, 
we derive the gas distribution in a halo.
Now the hydrostatic equilibrium gives
%%%%%
\begin{eqnarray}
{(1+\epsilon)\over \rho_g}\frac{dP_g}{dr}=-\frac{1}{\mpl}\left[\frac{d\phi_G}{dr}+\beta\frac{d\phi}{dr}\right],
\label{hydrostaticeq}
\end{eqnarray}
%%%%%
where $\rho_g$ and $P_g$ are the gas density and the pressure, respectively, 
and $\epsilon=0$ unless explicitly stated otherwise.
Here $\phi_G$ denotes the gravitational potential, given by solving 
the gravitational Poisson equation, $\triangle \phi_G=\rho/(2\mpl)$.
For the generalized NFW density profile, we find the solution 
%\begin{eqnarray}
$\phi_G(x)=%\left\{
%\begin{array}{ll}
\phi_0 [{1-(1+x)^{2-b}}/{(b-2)x}]$, %& {\rm for}~b\neq 2,\\
%\phi_0 \dfrac{\ln(1+x)}{x}, & {\rm for}~b=2,
%\end{array}
%\right.
%\end{eqnarray}
where we define $\phi_0=-\rho_sr_s^2/2\mpl(b-1)$.
%In this letter, for simplicity, we adopt the following relation  
%suggested by N-body simulations \cite{c-M},
%$c=6\left[{\mvir}/{10^{14}h^{-1}M_{\odot}}\right]^{-1/5}$.
%Hence, a halo is specified by the virial mass $M_{\rm vir}$. 
%%%%
%Then, the chameleon force on a test particle with unit mass is given by
%combining the gravitational force and the chameleon force
%%%%
%\begin{eqnarray}
%f(r)=-\frac{1}{\mpl}\left[\frac{d\phi_G}{dr}+\beta\frac{d\phi}{dr}\right],
%f(r)=-\frac{G\mvir}{m(c)}
%\frac{m^\prime(r/r_s)}{r^2}
%\end{eqnarray}
%with
%\begin{eqnarray}
%m^\prime(x)=
%\left\{
%\begin{array}{ll}
%+\dfrac{2\beta^2\phi_s}{B}
%\dfrac{x(1+3x)\left[x(1+x)^2\right]^{1/(1+n)}}{(1+n)(1+x)} & ~ \\
%~~~~~~~~-\dfrac{x}{1+x}+\ln(1+x), ~~~~~~~x<x_c, &~ \\
%
%~~~~~~~~~~~~~~~~~~~~~~~~~~~~~~~~~& ~~\\
%\dfrac{2\beta^2 C}{B}-
%\dfrac{(1+2\beta^2)x}{(1+x)}+(1+2\beta^2)\ln(1+x), &~ \\
%~~~~~~~~~~~~~~~~~~~~~~~~~~~~~~~~~~~~~~~~~~~x>x_c, & 
%\end{array}
%\right.
%\nonumber
%\end{eqnarray}
We assume that the gas obeys the polytropic equation of 
state $P_g\propto \rho_gT_g\propto \rho_g^\gamma$ with the 
polytropic index $\gamma$ and the gas temperature $T_g$.
Introducing the function $y_g(x)$ by $\rho_g(x)=\rho_g(0)y_g(x)$,
$P_g(x)=P_g(0)y_g^{\gamma}(x)$, and $T_g(x)=T_g(0)y_g^{\gamma-1}(x)$, 
we obtain the solution
%%%%%
\begin{align}\label{yg}
&y_{g}(x) 
=\biggl[1 -\frac{\mu m_p}{kT_g(0)\mpl(1+\epsilon)}\dfrac{\gamma-1}{\gamma}
\nonumber\\
&~~~~~~~~
\times(\phi_G(x)-\phi_G(0)+\beta\phi(x)-\beta\phi(0))
\biggr]^{1/(\gamma-1)}
%\nonumber\\
\\
\label{exyg}
&=\left\{
\begin{array}{l}
\biggl[
1-A\biggl(1+\dfrac{(1+x)^{2-b}-1}{(b-2)x}
\\~~%~~~~~~~~~~~~~~~
-\displaystyle{\beta \phi_s\over \phi_0}[x(1+x)^b]^{1/(n+1)}\biggr)
\biggr]^{1/(\gamma-1)},
%\\~~~~~~~~~~~~~~~~~~~~~~~~~~~~~~~~~~~~~~~~
~~{\rm for}~x<x_c,
\\
\biggl[1-A
\biggl(1+(1+2\beta^2)\dfrac{(1+x)^{2-b}-1}{(b-2)x}
\\~~~~~~~~~~%~~~~~~~~~~~~~~~~~~~~~~~
-\displaystyle{\beta\over \phi_0}
\biggl(\phi_\infty-{C\over x}\biggr)\biggr)
\biggr]^{1/(\gamma-1)},
%\\~~~~~~~~~~~~~~~~~~~~~~~~~~~~~~~~~~~~~~~
~~~{\rm for}~x>x_c,
\end{array}
\right.
%\nonumber\\
\end{align}
%y_{g}(x) 
%=\left[1 -3\eta^{-1}(0)\dfrac{\gamma-1}{\gamma}\dfrac{c}{m(c)}\displaystyle\int_{0}^x\dfrac{m^\prime(u)}{u^2}du\right]^{1/(\gamma-1)},
%%%%%
%where
%%%%%
%\begin{eqnarray}
%m^\prime(x)=\left\{
%\begin{array}{ll}
%m^\prime_{\rm int}(x), & \quad x\leq x_c,\\
%m^\prime_{\rm ext}(x), & \quad x> x_c,
%\end{array}
%\right.
%\end{eqnarray}
%%%%%
%and
%%%%%
where we have defined
$A = -\mu m_p\phi_0(\gamma-1)/kT_g(0)\mpl(1+\epsilon)\gamma$,
$k$ is the Boltzmann constant, and $\mu m_p$ represents the mean molecular mass. 
We determine the parameter $\gamma$ by Expression (17) 
in Ref.~\cite{gamma-eta}. Our conclusions are not altered qualitatively
for the assumption on $\gamma$ within $1.1\leq \gamma \leq 1.3$.

%%%%%%%%%%%%%%%%%%%%%%%%%%%%%%
\begin{figure}[btp]
\includegraphics[width=70mm]{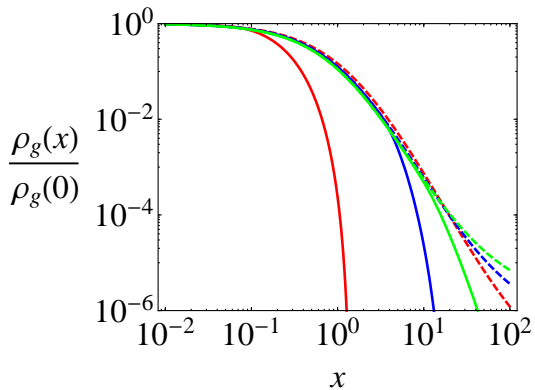}
\caption{
Gas density profile as a function of the radius $r/r_s$.
The solid (dashed) curves are with (without) the chameleon force,
with virial masses $\mvir=4\times 10^{14}M_{\odot}$ (green light curve), 
$10^{14}M_{\odot}$ (blue dark curve), and  $\mvir=10^{13}M_{\odot}$ 
(red curve), from top to bottom, respectively.
Here we have adopted $\beta=1$,~$n=1.8\times 10^{-5}$,  
$\Lambda=2.4\times 10^{-3}~{\rm eV}$, and $\phi_{\infty}=1.1\times 10^{-5}\mpl$.
}
\label{fig:density}
\end{figure}
%%%%%%%%%%%%%%%%%%%%%%%%%%%%%%

%The integration in expression (\ref{exyg}) yields
%%%%%
%\begin{eqnarray}
%&&\displaystyle\int_{0}^x\dfrac{m^\prime(u)}{u^2}du=
%\nonumber\\
%&&~~~~~~\left\{
%\begin{array}{l}
%\\
%1-\displaystyle{\ln(1+x)\over x}+\displaystyle{2\beta^2 \phi_s\over B}[x(1+x)^2]^{1/(n+1)}, \nonumber
%\\
%~~~~~~~~~~~~~~~~~~~~~~~~~~~~~~~~~~~~~~~~~~~x<x_c
%\\
%1+\displaystyle{2\beta^2\over B}\phi_\infty
%-\displaystyle{2\beta^2 C\over B}{1\over x}-(1+2\beta^2)\displaystyle{\ln (1+x)\over x}, 
%\nonumber
%\\
%~~~~~~~~~~~~~~~~~~~~~~~~~~~~~~~~~~~~~~~~~~~x>x_c,
%\end{array}
%\right.
%\end{eqnarray}
%%%%%

Figure \ref{fig:density} shows the gas density profiles, comparing the case with the
chameleon force (solid curves) and the case of Newtonian gravity (dashed curves),
adopting virial masses of $\mvir=10^{13}M_\odot$, $10^{14}M_\odot$, and  
$4\times 10^{14}M_\odot$, from top to bottom, respectively.
The gas density decreases rapidly in the outer region (see the solid curves), 
where the chameleon force is influential. 
For the large mass cluster, the chameleon mechanism works out to large radii, 
because the density of dark matter is high enough even outside the halo. 
On the other hand, for the small mass cluster, the chameleon mechanism works 
only at small radii, because the dark matter density is high only in the
central region. 
Because the chameleon force is an attractive force, a larger pressure 
gradient is necessary for balancing between them.
This makes the gas distribution compact. 
This feature is more significant for the smaller-mass halo.

%%%%%%%%%%%%%%%%%%%%%%%%%%%%%%
\begin{figure}[t]
\includegraphics[width=60mm]{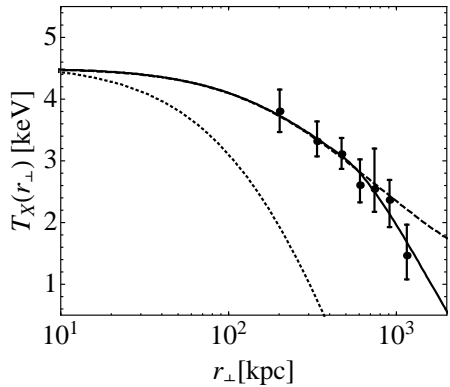}
\caption{
Temperature profiles as a function of the radius $r_\perp$. 
The points with error bars show the observation data
of the filament direction \cite{data}.
The curves show our theoretical model. 
The solid curve adopts the best-fit parameters 
$(\phi_\infty, M_{\rm vir}, c, T_g(0))=
(5.4\times 10^{-5}\mpl, 5.1\times 10^{14}M_\odot, 5.8, 4.9{\rm keV})$. 
%but the red solid curve assumes 
%$(6.0\times 10^{-5}\mpl, 4.6\times 10^{14}~M_\odot, 4.09, 3.89~{\rm keV})$. 
The dashed curve adopts $\phi_\infty=2\times 10^{-5} \mpl$, while the
dotted curve adopts $\phi_\infty=1.3\times 10^{-4} \mpl$, where the other
parameters are the same as those of the solid curve. 
%The dashed curve is almost the same as that in the Newton limit, 
%while the dotted curve is almost the same as that in the limit of 
%modified gravity.
Here we have fixed $\beta=1$ and $b=2$. 
}
\label{fig:profile}
\end{figure}
%%%%%%%%%%%%%%%%%%%%%%%%%%%%%%
%-----------------------------------------------------------
%\section 
%{\it Observational Signatures}~
%-----------------------------------------------------------
Using this characteristic feature, let us consider a constraint on the 
chameleon gravity model. 
To this end, we consider x-ray observations of a cluster of galaxies. 
%which is produced by the bremsstrahlung of the ionized gas. 
% and the Sunyaev-Zel'dovich (SZ) 
%effect in the cosmic microwave background measurement. 
%The X-ray emission from clusters of galaxies 
%For example, the X-ray surface brightness is computed by $I(r_\perp)
%\propto\int_{-\infty}^\infty \lambda_c(T_g)\rho_g^2(\sqrt{r_\perp^2+z^2})dz$, 
Because of the steep drop of the gas density in the presence of the
chameleon force, a similar drop in the x-ray surface 
brightness may appear in the outer region. 
In the present paper, we compare the x-ray temperature profile from the data 
reported from Suzaku observations of the Hydra A cluster out to the virial 
radius~\cite{data}.
The Hydra A cluster is a medium-sized cluster located at a distance of $230$ Mpc. 
Two different fields are observed in Ref.~\cite{data}.
One is the northwest offset from the x-ray 
peak of the cluster, and the other is the northeast offset. 
The former and latter fields are called the filament and void, respectively, 
because each field continues into the filament and void structures. 
In Fig.~\ref{fig:profile}, the points with error bars show the 
data of the filament direction in Ref.~\cite{data}. 

The curves in Fig.~\ref{fig:profile} show our theoretical model of the x-ray surface 
brightness temperature, computed with the formula
\begin{eqnarray}\label{temperature}
T_X(r_\perp)=\frac
{\int\lambda_c(T_g)\rho_g^2(\sqrt{r_\perp^2+z^2})T_g(\sqrt{r_\perp^2+z^2})dz}
{\int\lambda_c(T_g)\rho_g^2(\sqrt{r_\perp^2+z^2})dz},
\end{eqnarray}
where $\lambda_c(T_g)$ is the cooling function, for which we have assumed 
$\lambda_c(T_g)\propto T_g^{1/2}$ (e.g., Ref.~\cite{coolingfunction}), 
and $r_\perp$ is the radial coordinate perpendicular to the line-of-sight direction.
The solid curve is the best-fit curve, whose parameters are noted in 
the caption. 
The dashed curve and the dotted curve adopt 
$\phi_\infty=2\times 10^{-5} \mpl$ and $1.3\times 10^{-4} \mpl$, 
respectively, where the other parameters are the same as those for 
the solid curve.

%%%%%%%%%%%%%%%%%%%%%%%%%%%%%%
\begin{figure}[t]
\includegraphics[width=70mm]{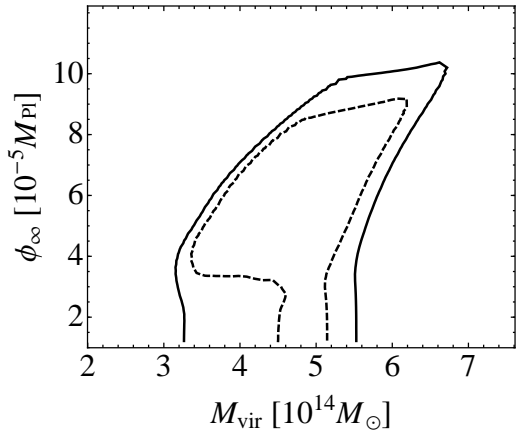}
\caption{
The contours of $\Delta\chi^2$ on the parameter plane $\phi_\infty$-
$M_{\rm vir}$. Here we have fixed $\beta=1$ and $b=2$, but $c$ and $T_g(0)$
are varied as fitting parameters. 
The contour levels of the inner dashed curve and the outer solid curve are
$\Delta\chi^2=2.7$ and $6.6$, respectively.
%Here we adopted $n=1.8\times 10^{-5}$ and $\Lambda=2.4\times 10^{-3}~{\rm eV}$, 
%but the results don't depend on the values of $n$ and $\Lambda$.
}
\label{fig:chisquare1}
\end{figure}
%%%%%%%%%%%%%%%%%%%%%%%%%%%%%%

The dotted curve, the solid curve, and the dashed curve in 
Fig.~\ref{fig:profile} represent the characteristic
curves which appear when we vary $\phi_\infty$ from a sufficiently 
large value to a smaller one. 
First, the dotted curve represents the limit of the modified gravity. 
Namely, for the large value of $\phi_\infty\geq B$, $x_c$ becomes negative from 
Eq.~(\ref{connection2}). This means that there appears no interior region in a halo 
where the chameleon mechanism works to recover Newtonian gravity. 
Thus, for the case $\phi_\infty\geq B$, we have $\phi(x)=\phi_{\rm out}(x)$ for the 
entire region, and therefore the solution Eq.~(\ref{yg}) should be replaced with
\begin{align}\label{yg2}
y_g(x)=\biggl[
1-A(1+2\beta^2)\biggl(1+\dfrac{(1+x)^{2-b}-1}{(b-2)x}\biggr)
\biggr]^{1/(\gamma-1)}.
\nonumber\\
\end{align}
On the other hand, the dashed curve is the same as the limit of Newtonian gravity. 
For a small value of $\phi_\infty$, we have a large value of $x_c$ from Eq.~(\ref{connection2}). 
This means that the chameleon force is influential only at very large radii. 
Note that the interior solution $y_g(x)$, Eq.~(\ref{exyg}) for $x<x_c$, 
can be approximated by taking the limit of $\beta\to 0$ in Eq.~(\ref{yg2}),
because $\phi_s$ takes a very small value to screen the scalar field where 
the chameleon mechanism works.
In summary, the dotted curve and the dashed curve are the two opposite limits, and
our theoretical curve is restricted by these two limits. Note that the limit of the 
modified gravity Eq.~(\ref{yg2}) depends on the coupling constant $\beta$. 

\begin{table}[b]
\begin{tabular}{c|c|c}
\hline
&\multicolumn{2}{|c}{Upper limit for $\phi_\infty$ in unit of $[\mpl]$}\\
\cline{2-3}
& Filament & Void\\
\hline
~~~~~~$b=1.7$~~~~~~&~~~~~~$1.4\times 10^{-4}$~~~~~~&~~~~~~$0.9\times 10^{-4}$~~~~~~\\
\hline
~~~$b=2.0$~~~&~~~$1.0\times 10^{-4}$~~~&~~~$0.8\times 10^{-4}$~~~\\
\hline
~~~$b=2.5$~~~&~~~$0.8\times 10^{-4}$~~~&~~~$0.6\times 10^{-4}$~~~\\
\hline
\end{tabular}
\caption{Upper bounds of $\phi_\infty$ at the $2$-sigma level
for different values of $b$ and the data for the filament and void directions.
Here we have fixed $\beta=1$.}
\label{constraint}
\end{table}

Figure \ref{fig:chisquare1} shows the contours of $\Delta\chi^2$ on the parameter plane for
$\phi_\infty$ and $M_{\rm vir}$,  where $\chi^2$ is simply defined 
by $\chi^2=\sum_{i=1}^7(T_X(r_{\perp,i})-T_i^{\rm obs.})^2/(\Delta T_i^{\rm obs.})^2$, 
where $T_i^{\rm obs.}$ and $\Delta T_i^{\rm obs.}$ are the observed data and 
the error of the filament direction, respectively, and 
$T_X(r_{\perp,i})$ is our theoretical model. 
Here, we have fixed $\beta=1$ and $b=2$, but the parameters $c$ and $T_g(0)$ are
varied so as to minimize $\chi^2$ within the range $3\leq c\leq 10$ and $T^*_0/\alpha\leq T_g(0)
\leq T^*_0\alpha$ with $\alpha=1.1$, where $T^*_0$ is given by Eq.~(19) in Ref.~\cite{gamma-eta}.
When taking $T_g(0)$ as a completely free parameter, it is difficult to obtain
a useful constraint from the present data due to the degeneracy between $T_g(0)$ and $\mvir$.
The minimum value of $\chi^2$ is $1.0$, where the 
number of degrees of freedom is $3$. 
The behavior of the contour is explained by the fact that
the theoretical curve approaches that of Newtonian gravity as 
$\phi_\infty$ becomes small and that the steep drop becomes significant as 
$\phi_\infty$ increases. 
Figure~\ref{fig:chisquare1} gives an upper 
bound of $\phi_\infty<10^{-4}\mpl$ at the $2$-sigma %statistical 
level for the case 
$b=2$ and $\beta=1$. 
We obtain a similar upper bound of $\phi_\infty$ for 
different values of $b$, which are 
summarized in Table~\ref{constraint}. 
The upper bound of $\phi_\infty$ becomes larger for smaller $b$, but we may 
conclude that the results do not significantly depend on $b$. 
Table~\ref{constraint} includes the results with the void direction. 
The upper bound of $\phi_\infty$ depends on the data, i.e., the filament direction
and the void direction; however, our conclusion does not alter significantly. 

%For smaller $b$, the temperature profile bocomes {\bloose ???
%Then $\phi_\infty$ is allowed to a somewhat high value
%compared with the case that $b$ is high value.

So far, we have considered the case $\beta=1$; let us now discuss the case
$\beta=1/\sqrt{6}$, which corresponds to an $f(R)$ model \cite{starobinsky}. 
In this case, we could not obtain a useful constraint on $\phi_\infty$, 
which is explained as follows.
%which is related with the parameter $f_{R0}$ in the $f(R)$ model. 
%by $f_{R0}=-\sqrt{2/3}{\phi_\infty/\mpl}$. 
The theoretical density profile is limited by two characteristic 
curves, Eq.~(\ref{yg2}) and Eq.~(\ref{yg2}) with $\beta=0$. 
When $\beta$ is small, the difference between these two characteristic curves is small, 
because the drop of the gas density is not steep. This is the reason why no 
useful constraint on the $f(R)$ model was obtained from the present x-ray data here. 
%In order to obtain a useful constraint, we need 
%a data with a smaller mass cluster. 

%-----------------------------------------------------------
%\section 
%{\it Conclusions}~
%-----------------------------------------------------------
In summary, under the assumption of hydrostatic equilibrium between the gas 
pressure gradient and the gravitational and chameleon forces, we derived the 
gas density profile in an analytic manner. Here we assumed the polytropic equation 
of state for the gas and the generalized NFW density profile for the dark matter distribution. 
The chameleon force may give rise to a steep drop in the gas distribution in the outer region of a halo. 
This feature is more significant when the mass of a halo is small and $\beta$ and $\phi_\infty$
are large. The gas density profile depends on $\beta$ and $\phi_\infty$, 
but it does not depend on $n$ and $\Lambda$. 
This provides us with an opportunity to constrain $\beta$ and $\phi_\infty$ by comparison with
observations. We demonstrated a constraint on $\phi_\infty$ in the chameleon gravity 
model, using the data of the temperature profile from the x-ray observation \cite{data}.
We obtained a useful upper bound of $\phi_\infty<10^{-4}\mpl$ in the case $\beta=1$~\footnote{
When we take the environment effect into account, this bound might be understood as the bound
around the Hydra A cluster.}; however,
no useful constraint was obtained in the case $\beta=1/\sqrt{6}$, which corresponds to an $f(R)$ model. 
In order to obtain a useful constraint, observations of the outer region of a smaller 
mass cluster are more advantageous. Furthermore, a combination with other observations 
like the weak lensing measurements might improve the constraint. 
In our investigation, the assumption of the hydrostatic equilibrium of hot gas might be crucial. 
To estimate the effect of deviation from it, we obtained similar constraints
by adopting the nonzero values of $\epsilon=\pm0.5$ in Eq.~(\ref{hydrostaticeq}). 
The upper bound of $\phi_\infty$ changes from $10^{-4}\mpl$ for $\epsilon=0$ to 
$0.6\times10^{-4}\mpl$ and $2.1\times10^{-4}\mpl$ for $\epsilon=0.5$ and $-0.5$, respectively.
Thus, the assumption of hydrostatic equilibrium is crucial to the constraint, but 
we may obtain a useful constraint if we can model the state of the gas correctly.  
Further study is necessary for this problem.
Finally, we assumed spherical symmetry for a halo, an assumption whose validity must be checked when
comparing with observational data. In the present paper, the results in Table~\ref{constraint} 
do not depend on the filament direction and the void direction significantly, which 
suggests the validity of our assumption.

\vspace{0.mm}
{\it Acknowledgments}~
We thank R.~Kimura and T.~Narikawa for fruitful discussions.
We also thank S.~Nishino, Y.~Fukazawa, T.~Kobayashi, K.~Koyama, R.~C.~Nichol, 
A.~Taruya, and Y.~Suto for useful conversation related to the topic in the present paper.
This work was supported by a Grant-in-Aid for Scientific
Research from the Japanese Ministry of Education,
Culture, Sports, Science, and Technology (Grants No.~21540270, No.~21244033).
%and JSPS Core-to-Core Program
%``International Research Network for Dark Energy''.

\end{document}